\documentclass[nofootinbib,aps,11pt,amsmath,amssymb,a4paper,prd,preprintnumbers]{revtex4-1}

\usepackage[american]{babel}
\usepackage[utf8]{inputenc}
\usepackage[T1]{fontenc}

\usepackage{graphicx}
\usepackage{color}

\usepackage[caption=false]{subfig}

\usepackage{slashed}

\usepackage{units}

\usepackage{hyperref}
\hypersetup{
   pdfnewwindow=true,    
   colorlinks=true,      
   linkcolor=blue,       
   citecolor=blue,       
   filecolor=black,       
   urlcolor=blue         
}

\begin{document}

\title{\Large {\bf{New Forces and the 750 GeV Resonance}}}
\author{Michael Duerr,$^{1}$ Pavel Fileviez P\'erez,$^{2}$ Juri Smirnov$^{2}$}        
\affiliation{\vspace{3ex} $^{1}$DESY, Notkestra\ss{}e 85, D-22607 Hamburg, Germany \\
$^{2}$Particle and Astroparticle Physics Division, Max-Planck-Institut f\"ur Kernphysik, \\ Saupfercheckweg 1, 69117 Heidelberg, Germany } 
\email[E-mail addresses: ]{michael.duerr@desy.de}
\email{fileviez@mpi-hd.mpg.de}
\email{juri.smirnov@mpi-hd.mpg.de}



%
\begin{abstract}  
Recently, the ATLAS and CMS collaborations have pointed out the possible existence of a new resonance with a mass around $\unit[750]{GeV}$. 
We investigate the possibility to identify this new resonance with a spin zero field responsible for the breaking of a new gauge symmetry. 
We focus on a simple theory where the baryon number is a local symmetry spontaneously broken at the low scale.
In this context new vector-like quarks are needed to cancel all baryonic anomalies and define the production mechanism and decays of the new Higgs at the LHC.  
Assuming the existence of the new Higgs with a mass of $\unit[750]{GeV}$ at the LHC we find an upper bound on the symmetry breaking scale. 
Therefore, one expects that a new force associated with baryon number could be discovered at the LHC.
\end{abstract}

\preprint{DESY 16-070}

\maketitle



\section{Introduction}
The Large Hadron Collider~(LHC) could soon find signatures of new physics and the Standard Model~(SM) will be part of the past of particle physics.
Recently, both the ATLAS and CMS collaborations have pointed out the possible existence of a new resonance due to an excess in the di-photon channel~\cite{ATLAS:diphotonDec2015,CMS:2015dxe,ATLAS:diphotonMar2016,CMS:2016owr}.
With $\unit[3.2]{fb}^{-1}$ of data at $\sqrt{s} = \unit[13]{TeV}$, the ATLAS collaboration claims an excess in the di-photon channel around a di-photon invariant mass of $M_{\gamma \gamma} \approx \unit[750]{GeV}$ with a local significance of $3.9\, \sigma$ ($2.0\, \sigma$ including the look-elsewhere effect)~\cite{ATLAS:diphotonMar2016}. The CMS collaboration also reports an excess with a local statistical significance of $3.4 \,\sigma$ ($1.6 \, \sigma$ including the look-elsewhere effect) in the same invariant mass region~\cite{CMS:2016owr}. From the conservative point of view it may be too early to speculate 
about the existence of a new particle in nature. Nevertheless, it is interesting to know whether this excess can be explained in a well-known extension of the Standard Model 
or new classes of theories for physics beyond the Standard Model have to be build.

One of the simplest toy models proposed in the literature corresponds to the case of a new SM singlet scalar field $S$ and vector-like pairs of fermions $F_L$ and $F_R$ with electric charge in a non-trivial representation of QCD. The relevant Lagrangian of this model is given by
\begin{equation}
- {\mathcal{ L}} \supset m_{F} \overline{F_L} F_R \ + \ \lambda_{F} S  \overline{F_L} F_R \ + \ \textrm{h.c.},
\end{equation} 
where $m_{F}$ is the vector-like mass and $\lambda_{F}$ is the Yukawa coupling between 
the new Higgs and the vector-like fermions. Since the new fermions live in a non-trivial representation of QCD, one can have 
the single production of $S$ through gluon fusion, and the decay of into two photons is possible because the new fermions carry electric charge.
This toy model has been studied by several groups, see Refs.~\cite{Franceschini:2015kwy,Ellis:2015oso,Gupta:2015zzs,McDermott:2015sck,Dutta:2015wqh,Kobakhidze:2015ldh,Chao:2015ttq,Curtin:2015jcv,Falkowski:2015swt,Benbrik:2015fyz,Chao:2015nsm,
Chang:2015bzc,Feng:2015wil,Boucenna:2015pav,Hernandez:2015ywg,Pelaggi:2015knk,Altmannshofer:2015xfo,Cheung:2015cug,An:2015cgp,Dev:2015vjd,Ko:2016lai,Chao:2016mtn}.

The above toy model is very naive and one should look for a UV completion of the Standard Model where one can understand the need for these vector-like quarks. If one considers a simple $U(1)^\prime$ theory one can motivate the existence of vector-like quarks. This type of model has been studied in Refs.~\cite{Chao:2015nsm,Chang:2015bzc} as an attempt to explain the di-photon excess.

In this article we investigate the possibility to explain the di-photon excess in a theory for local baryon number~\cite{FileviezPerez:2010gw,FileviezPerez:2011pt,Duerr:2013dza,Perez:2014qfa,Perez:2015rza}, where one needs to introduce vector-like quarks to cancel all baryonic anomalies~\cite{FileviezPerez:2010gw}. We study the decays of the new Higgs boson that is responsible for symmetry breaking and show that this Higgs can give 
rise to the di-photon signatures reported by the ATLAS and CMS collaborations. Assuming the existence of the di-photon excess with invariant mass around 
$\unit[750]{GeV}$, we find an upper bound on the symmetry breaking scale. Therefore, if this excess is real and the relevant theory corresponds to the case where baryon number is spontaneously broken, one should find a new force associated with baryon number at the Large Hadron Collider.

\section{A Theory for the Di-Photon Excess}
\label{sec:localbaryonnumber}
One can define a simple theory where baryon number is a local symmetry spontaneously broken at the low scale~\cite{FileviezPerez:2010gw,FileviezPerez:2011pt,Duerr:2013dza,Perez:2014qfa,Perez:2015rza}.
In Ref.~\cite{FileviezPerez:2011pt} it was shown that using vector-like quarks one can cancel all baryonic anomalies. 
The fermions needed for anomaly cancellation and their quantum numbers under the new gauge group 
$$SU(3)_C \otimes SU(2)_L \otimes U(1)_Y \otimes U(1)_B$$
are given by
\begin{align}
\Psi_L &\sim (3,2,Y_1,B_1), \ \eta_R \sim (3,1,Y_2,B_1), \ \chi_R \sim (3,1,Y_3, B_1), \nonumber \\ 
\Psi_R &\sim (3,2,Y_1,B_2), \ \eta_L \sim (3,1,Y_2,B_2), \ \chi_L \sim (3,1,Y_3, B_2).
\end{align}
The hypercharges and baryon numbers of these fields can be fixed by the conditions from anomaly cancellation as follows. Cancellation of the $SU(2)_L^2 \otimes U(1)_B$ anomaly requires
\begin{equation}
 B_1-B_2=- \frac{1}{n_f},
\end{equation}
where $n_f$ is the number of copies of the vector-like quarks. Using this condition, $U(1)_B \otimes U(1)_Y^2$ anomaly cancellation requires
\begin{equation}
\label{eq:Hypercharge}
Y_2^2+Y_3^2-2Y_1^2 = \frac{1}{2},
\end{equation}
while the cancellation of the $U(1)_Y \otimes U(1)_B^2$ anomaly for $B_1 \neq - B_2$ implies
\begin{equation}
Y_2 +Y_3 -2 Y_1 = 0.
\end{equation}
Both conditions on the hypercharges are satisfied if
\begin{equation}\label{eq:hyperchargeSolution}
Y_2=Y_1 \mp \frac{1}{2}  \text{ and } Y_3= Y_1 \pm \frac{1}{2}.
\end{equation}
This relation allows to write the Yukawa couplings for the new quarks with the SM Higgs, which are relevant for the decays of the heavy new quarks. 

The relevant Lagrangian for our discussion is then given by
\begin{align}
\label{eq:Lagrangian}
- \mathcal{L}_Y^\prime &= h_1 \overline{\Psi_L} \tilde{H} \eta_R + h_2 \overline{\Psi_L} H \chi_R + h_3 \overline{\Psi_R} \tilde{H} \eta_L + h_4 \overline{\Psi_R} H \chi_L \nonumber \\
& \quad + \lambda_\Psi \overline{\Psi_R} \Psi_L S_B + \lambda_\eta \overline{\eta_L} \eta_R S_B + \lambda_\chi \overline{\chi_L} \chi_R S_B \ + \  \text{h.c.},
\end{align}
where $S_B \sim (1,1,0,B_2-B_1)$ is the new Higgs breaking the local symmetry and generating masses for the vector-like quarks. If the baryon numbers $B_i \neq 1/3$, there is no mixing between the vector-like quarks and the SM quarks and thus no flavor violation at tree level.

The above conditions on the hypercharges have three solutions if we demand that at least one hypercharge is equal to a SM quark hypercharge, which we regard as necessary in order for the lightest new colored field to decay. The three solutions that can be in agreement with cosmology are 
\begin{equation}
\left( Y_1; Y_2; Y_3\right)\in \left\{ \left( \frac{1}{6} ; \frac{2}{3} ; -\frac{1}{3}\right) , \left(-\frac{5}{6} ; -\frac{4}{3} ; -\frac{1}{3}\right), \left( \frac{7}{6} ; \frac{5}{3} ; \frac{2}{3}\right)\right\}.
\end{equation}
The first solution was proposed in Ref.~\cite{FileviezPerez:2011pt}; see also Ref.~\cite{Duerr:2013dza}. 
In Ref.~\cite{FileviezPerez:2011pt} it was proposed that the vector-like quarks can decay 
into the SM quarks and a bosonic field $X$, which is automatically stable and a candidate for the cold dark matter in the Universe if it does not acquire a vacuum expectation value. 
This condition defines the three different scenarios that we study in detail in this article.

\begin{itemize}
\item Type I Scenario:

In this case the hypercharges for the new vector-like quarks are $Y_1=1/6$, $Y_2=2/3$ and $Y_3=-1/3$, and the relevant Lagrangian reads as
\begin{equation}
\mathcal{L}_I \supset  \mathcal{L}_Y^\prime  - \lambda_1 \overline{\Psi_R} Q_L X - \lambda_2 \overline{\eta_L} u_R X - \lambda_3 \overline{\chi_L}
 d_R X + \text{h.c.}
 \end{equation}   
Notice that in this case all the new quarks can decay into a SM quark and $X$. Therefore, they can give rise to signatures with two jets and missing energy at the LHC. 

\item Type II Scenario: 

This scenario is defined by the coupling of one of the vector-like quarks to $d_R$. Therefore, $Y_1=-5/6$, $Y_2=-4/3$ and $Y_3=-1/3$ and the interactions are given by
\begin{equation}
\mathcal{L}_{II} \supset \mathcal{L}_Y^\prime (H \leftrightarrow \tilde{H}) - \lambda_3 \overline{\chi_L} d_R X + \text{h.c.}
 \end{equation}   
\item Type III Scenario:

For $Y_1=7/6$, $Y_2=5/3$ and $Y_3=2/3$ the lightest new quark can decay into the $u_R$ quark and dark matter via
\begin{equation}
 \mathcal{L} \supset   \mathcal{L}_Y^\prime  - \lambda_3 \overline{\chi_L}
 u_R X + \text{h.c.}
 \end{equation}   
\end{itemize}
In the above equations $X\sim (1,1,0,B_2-1/3)$ is our dark matter candidate. 
The scalar potential will contain all possible terms between the SM Higgs $H$ and the new scalar fields $X$ and $S_B$. Here we list only the portal terms between the scalar fields:
\begin{equation}
V \supset  \lambda_{HB} S_B^\dagger S_B H^\dagger H + \lambda_{HX} X^\dagger X H^\dagger H +  \lambda_{BX} S_B^\dagger S_B X^\dagger X \ + \  \text{h.c.}
 \end{equation}
The mass matrix for the up-type vector-like quarks in the basis $U_L=\left( \Psi^u_L \ \eta_L \right)$ and $U_R=\left( \Psi^u_R \ \eta_R  \right)$ is given by
\begin{equation}
\mathcal{M_U}=\frac{1}{\sqrt{2}}\left( \begin{matrix}
\lambda_\Psi v_B  & h_3 v_0  \\
h_1^* v_0  & \lambda_\eta^* v_B^* 
\end{matrix} 
\right),
\end{equation}
while the mass matrix for the down-type vector-like quarks in the basis $D_L=\left(\Psi^d_L \ \chi_L  \right)$ and $D_R=\left( \Psi^d_R \ \chi_R  \right)$ reads as
\begin{equation}
\mathcal{M_D}= \frac{1}{\sqrt{2}}\left( \begin{matrix}
 \lambda_\Psi v_B & h_4 v_0  \\
h_2^* v_0  & \lambda_\chi^* v_B^*
\end{matrix} 
\right).
\end{equation}
Here we have used $S_B=\frac{1}{\sqrt{2}} (v_B + h_B)+ \frac{i}{\sqrt{2}} A_B$, where $v_B$ is the vacuum expectation value and $A_B$ is the Goldstone boson eaten by the leptophobic gauge boson.
Neglecting the off-diagonal terms in the mass matrices above  and defining the Dirac four component fields $U_i^T=(U_{L i}  \ U_{R i})$ and $D_{i}^T=( D_{L i} \ D_{R i})$, the Lagrangian can be written as
\begin{equation}
- \mathcal{L} \supset  \sum_{i=1}^2 \left( M_{U_i} \overline{U}_i U_i + \frac{M_{U_i}}{v_B} h_B \overline{U_i} U_i + M_{D_i} \overline{D}_i D_i + \frac{M_{D_i}}{v_B} h_B \overline{D_i} D_i  \right),
 \end{equation}   
where
\begin{equation}
M_\Psi = M_{U_1} =   M_{D_1}= \lambda_\Psi \frac{v_B}{\sqrt{2}}, \ \ M_\eta = M_{U_2}=\lambda_\eta \frac{v_B}{\sqrt{2}}, \ \ M_\chi = M_{D_2}=\lambda_\chi \frac{v_B}{\sqrt{2}}.
 \end{equation}  
Using these results and the Feynman rules in Appendix~\ref{app:feynman}, we are ready to study the production and decays of the Higgs $h_B$ responsible for symmetry breaking. 
For simplicity, we will assume that all the new quarks have the same mass $m_Q$.

\section{A \texorpdfstring{$\boldsymbol{750}$}{750} GeV Higgs and the Symmetry Breaking Scale}
\label{sec:lhc}
%
\subsection{Decays of \texorpdfstring{$\boldsymbol{h_B}$}{hB}}

If one has the term $H^\dagger H S_B^\dagger S_B$ in the scalar potential, the new physical Higgs $h_B$ will decay into all the SM quarks, the new quarks and the SM gauge bosons at tree level.
Then, the branching ratio into two photons will be small since the predictions will be similar to the SM Higgs decays. Therefore, in this article we will assume that this term in the potential is very 
small. Even if there is no symmetry protecting this term one can find this term at one-loop level using the interactions in Eq.~(\ref{eq:Lagrangian}). However, the $h_i$ couplings in Eq.~(\ref{eq:Lagrangian}) can be very small as well.
The mixing term can then be found at two-loop level where inside the loop one has the top quark and the new gauge boson, see Ref.~\cite{Ohmer:2015lxa} for a discussion.  

In addition one can have tree level decays to the bosonic dark matter $X$ and unavoidably one-loop induced decays into the SM gauge bosons will occur. Also, if $h_B$ is heavy, one can have decays into two leptophobic $Z_B$ gauge bosons at tree level. In order to understand the possibility to explain the di-photon excess we will assume that the mixing to the SM Higgs is suppressed and the new quarks and gauge boson are heavy in order to avoid any tree-level decays of the physical Higgs $h_B$. 

\begin{figure}[t]
\includegraphics[width=0.47\linewidth]{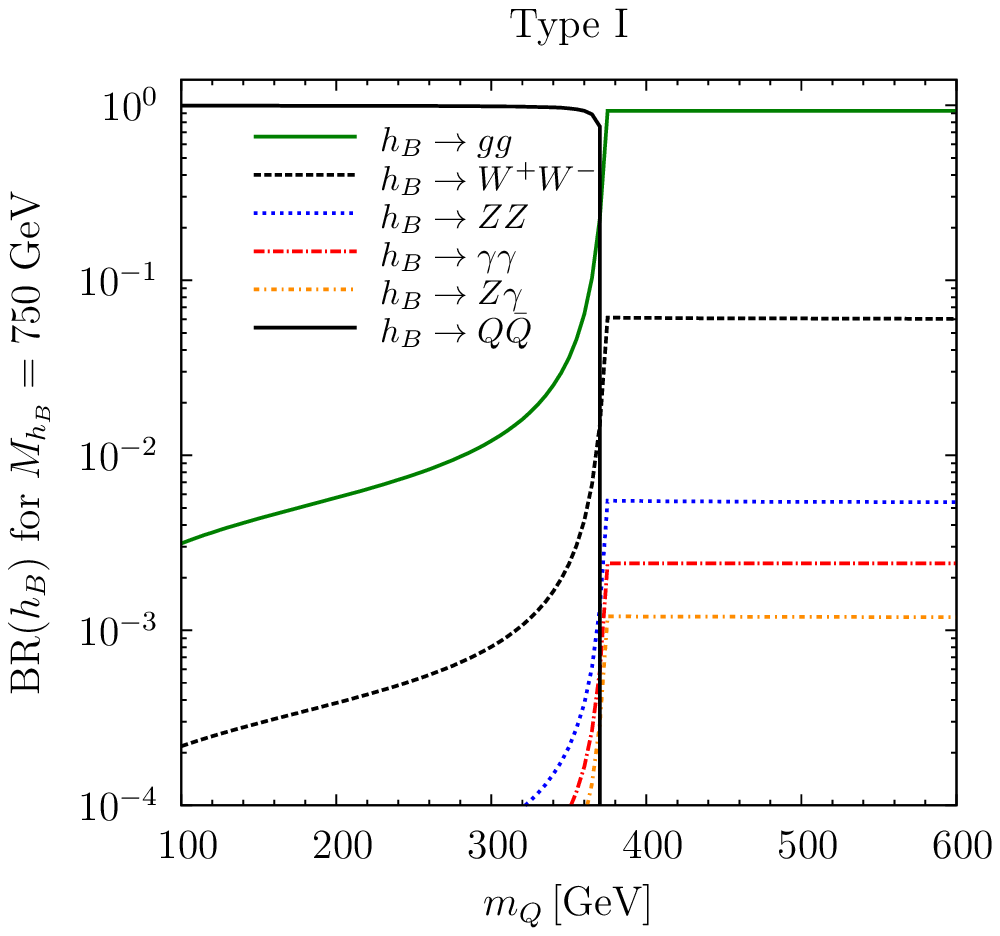}
\includegraphics[width=0.47\linewidth]{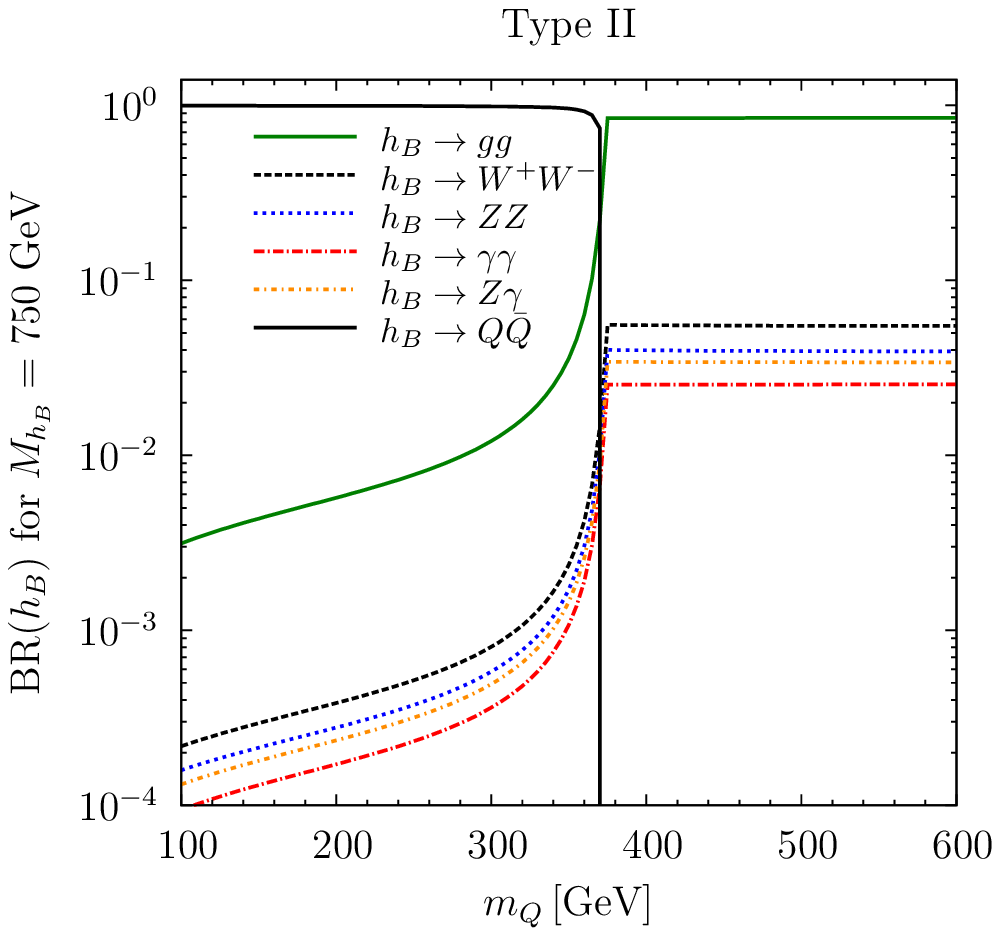}
\includegraphics[width=0.47\linewidth]{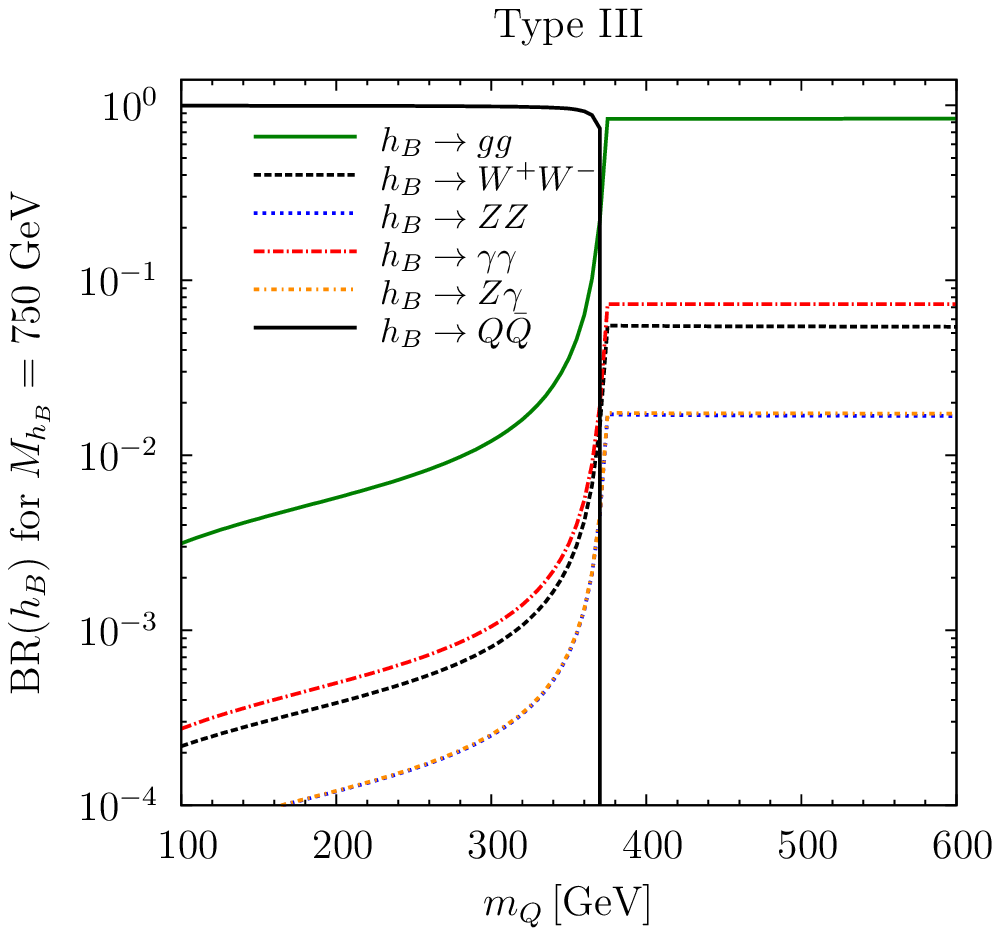}
\caption{Branching ratios of $h_B$ as a function of the common vector-like quark mass $m_Q$ for the Type I, II and II scenarios assuming that $M_{h_B} = \unit[750]{GeV}$. 
\label{fig:Brs}}
\end{figure}

In Fig.~\ref{fig:Brs} we show the branching ratios for $h_B$ as a function of the vector-like quark mass $m_Q$ that is assumed to be the same for all vector-like quarks. Notice that in the Type I and Type II scenarios the branching ratio into two photons is always smaller than the branching ratio into two W bosons. In the Type III 
scenario $\text{BR}(h_B \to \gamma \gamma) > \text{BR} (h_B \to WW)$ due to the fact that the electric charge of the vector-like quarks inside the loop is large. As we will see in the next section, in the Type III scenario one can easily explain the di-photon excess in agreement with all bounds.

\subsection{Production Mechanism at the LHC} 
The new Higgs $h_B$ that is responsible for symmetry breaking can be produced through gluon fusion due to its coupling to the new vector-like quarks.
The gluon fusion production cross section can be written as
\begin{equation}
\label{eq:SignalCrossSectionFull}
\sigma_{\gamma \gamma} \equiv \sigma (pp \to h_B) \times \text{BR}(h_B \to \gamma \gamma)= \frac{C_{gg}}{M_{h_B} \ \Gamma_{h_B} \ s} \Gamma (h_B \to gg) \Gamma (h_B \to \gamma \gamma),
\end{equation} 
where $\Gamma_{h_B}$ is the total decay width of $h_B$, $s$ is the square of the center-of-mass energy and $C_{gg}$ is the gluonic PDF contribution. 
Using MSTW2008 PDFs at NLO~\cite{Martin:2009iq}, $C_{gg}=2137$ for $\sqrt{s}= \unit[13]{TeV}$ and $M_{h_B} = \unit[750]{GeV}$. Working in
the most optimistic scenario where the decay is dominated by the decay into two gluons, one finds
\begin{equation}
\sigma (pp \to h_B) \times \text{BR}(h_B \to \gamma \gamma) \approx 6.6 \times 10^{3}\, \Gamma (h_B \to \gamma \gamma) \frac{\text{fb}}{\text{GeV}}.
\end{equation} 
This implies that independent of the model, the requirement for the photon decay width needs to be $\Gamma (h_B \to \gamma \gamma) \gtrsim \unit[0.5 \times 10^{-3}]{GeV}$ since the favored cross section value is $\sigma_{\gamma \gamma}= \unit[6\pm3]{fb}$ for CMS and $\sigma_{\gamma \gamma}= \unit[10\pm3]{fb}$ for ATLAS.
Now, in the limit of the mass of the new vector-like fermions being larger than $M_{h_B}$, one can write the decay width to two photons as 
\begin{equation}
\label{eq:PhotonWidth}
\Gamma (h_B \to \gamma \gamma) = \frac{9 \alpha^2 n_f^2 \,Q^4\,M_{h_B}^3}{144 \pi^3\,v_B^2} , 
\end{equation}
where $Q$ is the electric charge of a given vector-like quark inside the loop.

Using the lower bound on the decay width, naively $\Gamma (h_B \to \gamma \gamma) \gtrsim 0.5 \times 10^{-3}$, one can set an upper bound on the symmetry breaking scale using the above expression.
{\textit{This is a striking result which allows us to understand the testability of this model as a theory for the 750~GeV resonance. }}

With an upper bound on the symmetry scale one can also find perturbative upper bounds on the gauge boson mass and the masses of the vector-like fermions. These are given by
\begin{equation}
M_{Z_B} \leq 2 \sqrt{\pi} v_B^{\rm{max}} / n_f, \ { \rm{and}} \  M_\Psi, M_\eta, M_\chi \leq \sqrt{2 \pi} v_B^{\rm{max}}.
\end{equation}  
If the upper bound on the symmetry scale is not very large there is a hope to find a new force associated with baryon number and the new vector-like quarks needed for anomaly cancellation.
In the next section we perform a detailed numerical analysis to investigate accurately the values of the symmetry breaking scale and apply all relevant experimental constraints on the model.
We would like to comment that if we make the same study in a model where the SM leptons feel the new Abelian symmetry, one can show that it is not possible to explain the di-photon excess in agreement with the experimental bounds from the LEP2 experiment. 

\subsection{Upper Bound on the Symmetry Breaking Scale}
\label{sec:Pheno}

In this section we show the possibility to find an upper bound on the symmetry breaking scale if one assumes that $h_B$ has a mass around $\unit[750]{GeV}$ and the model explains the di-photon excess reported by ATLAS and CMS. In Table~\ref{tab:8TeVConstraints} we list the relevant experimental constraints from the $8$ TeV LHC data to understand if our results are in agreement with the previous experimental constraints. 

\begin{table}[b]
\caption{LHC $\sqrt{s}= \unit[8]{TeV}$ constraints as compiled in Ref.~\cite{Franceschini:2015kwy}. \label{tab:8TeVConstraints}}
\begin{center}
\begin{tabular}{ccc}
\hline
Observable & upper limit @ $\sqrt{s}=\unit[8]{TeV}$ \\
\hline \hline
$\sigma( p p \rightarrow \gamma \gamma)$ & $< \unit[1.5]{fb}$ \\
$\sigma( p p\rightarrow Z \gamma)$ & $< \unit[11]{fb}$ \\
$\sigma( p p \rightarrow Z Z)$ & $< \unit[12]{fb}$ \\
$\sigma( p p \rightarrow W W)$ & $< \unit[40]{fb}$ \\
$\sigma( p p \rightarrow j j)$ & $< \unit[2500]{fb}$ \\
$\sigma( p p \rightarrow \text{inv.})$ & $< \unit[800]{fb}$ \\
\hline
\end{tabular}
\end{center}
\end{table}   

\begin{figure}[t]
\includegraphics[width=0.6\linewidth]{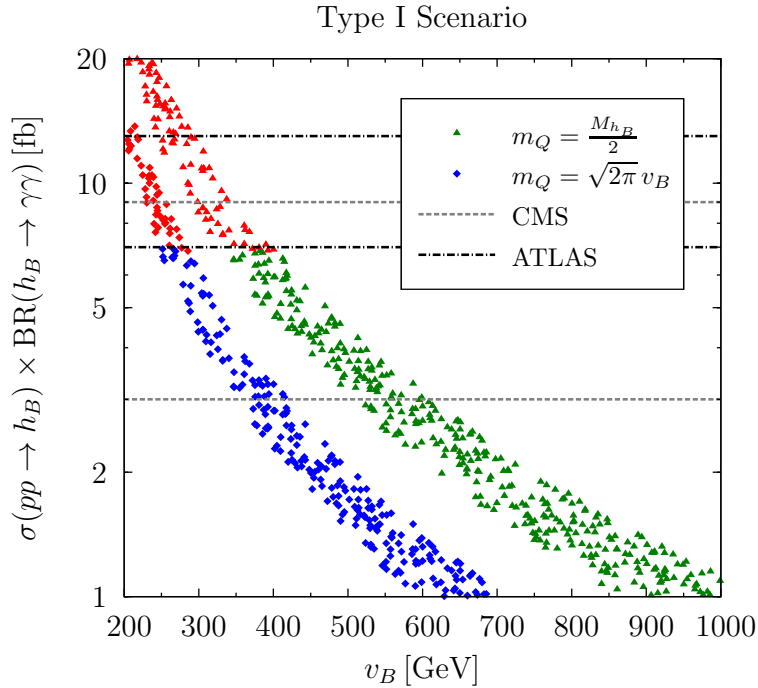}
\caption{Predictions for the cross section into two photons as a function of the symmetry breaking scale $v_B$ in the Type I scenario. The red points are excluded by $\unit[8]{TeV}$ data given in Table~\ref{tab:8TeVConstraints}.
The green (triangle) points correspond to the case when the vector-like quark masses are equal to half the Higgs mass, while the blue (square) points show the predictions when 
the vector-like masses are equal to the perturbative bound $m_{Q}=\sqrt{2 \pi} v_B$. The black dash-dotted horizontal lines show the favored range for ATLAS ($\sigma_{\gamma \gamma}=\unit[10\pm3]{fb}$), 
while the gray dashed horizontal lines show the preferred range for CMS ($\sigma_{\gamma \gamma}=\unit[6\pm3]{fb}$). The Higgs mass $M_{h_B}$ is varied between $700$ and $\unit[800]{GeV}$. Here we use the MSTW2008NLO PDFs~\cite{Martin:2009iq}, and Package-X for the one-loop calculations~\cite{Patel:2015tea}.
\label{fig:VEVRangesI}}
\end{figure}

In Fig.~\ref{fig:VEVRangesI}, we show the predictions for the cross section $\sigma (p p \to \gamma \gamma)$ versus the symmetry breaking scale $v_B$ in the Type I scenario. As one can see the allowed solutions are outside the range preferred by the ATLAS collaboration. 
Now, if we are conservative and use the allowed values by the CMS collaboration the upper bound on the symmetry breaking scale is $v_{B}^\text{max}=\unit[600]{GeV}$. Then, in the Type I scenario the perturbative upper bounds are $M_{Z_B} \leq \unit[2.1]{TeV}$ and $m_Q \leq \unit[1.5]{TeV}$. 

\begin{figure}[p]
 \includegraphics[width=0.6\linewidth]{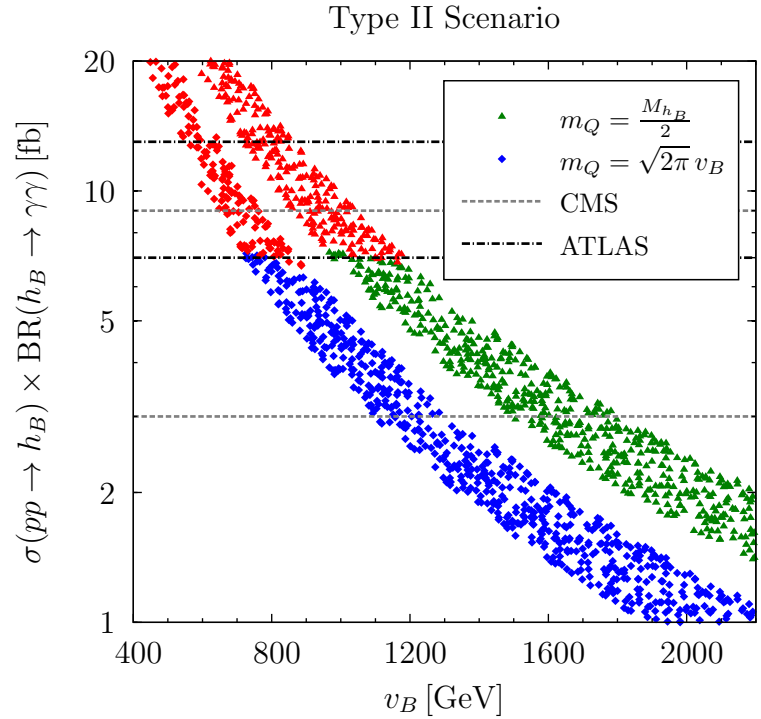}
\caption{The predictions for the Type II scenario. The input parameters are the same as in Fig.~\ref{fig:VEVRangesI}. \label{fig:VEVRangesII}}
\end{figure}

The predictions for $\sigma (p p \to \gamma \gamma)$ vs.\ $v_B$ for the Type II scenario are shown in Fig.~\ref{fig:VEVRangesII}. In this case one can find solutions for the cross sections which are in the overlap region for ATLAS and CMS results. Therefore, in this scenario one could have an explanation for the di-photon excess in agreement with the CMS and ATLAS. 
In this case the upper bound on the symmetry breaking scale is $v_B^\text{max}=\unit[1.8]{TeV}$, which gives us the perturbative bounds $M_{Z_B} \leq \unit[6.4]{TeV}$ and $m_Q \leq \unit[4.5]{TeV}$. 

\begin{figure}[p]
 \includegraphics[width=0.6\linewidth]{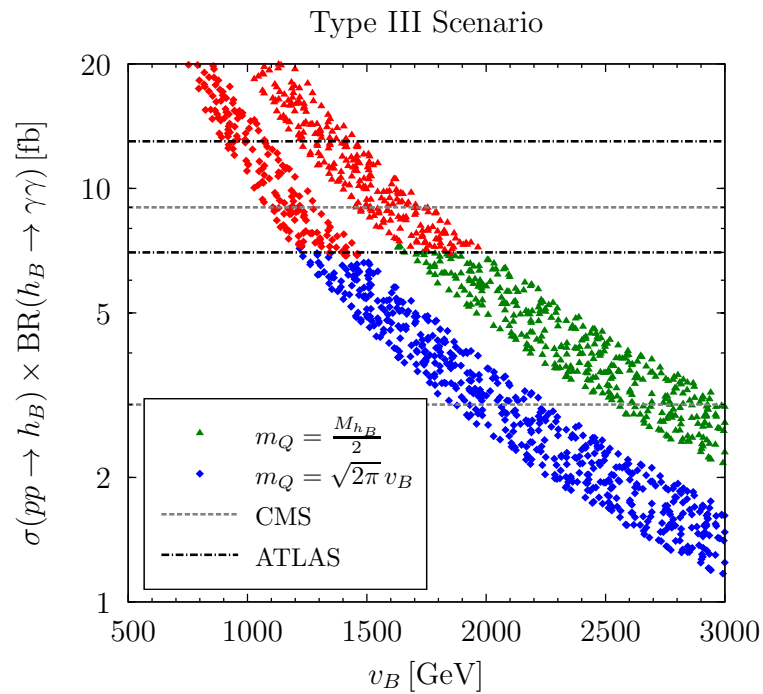}
\caption{The predictions for the Type III scenario. The input parameters are the same as in Fig.~\ref{fig:VEVRangesI}. \label{fig:VEVRangesIII}}
\end{figure}

In Fig.~\ref{fig:VEVRangesIII} we show the same results but for the Type III scenario. Notice that in this case the electric charge of the vector-like quarks inside the loop is larger and one can have a larger decay width for the decay into two photons. 
In this case the upper bounds are larger and are given by $v_{B}^\text{max}=\unit[3]{TeV}$, $M_{Z_B} \leq \unit[10.6]{TeV}$ and $m_Q \leq \unit[7.5]{TeV}$.

\begin{figure}[t]
\centering
\subfloat[]{
 \includegraphics[width=0.47\linewidth]{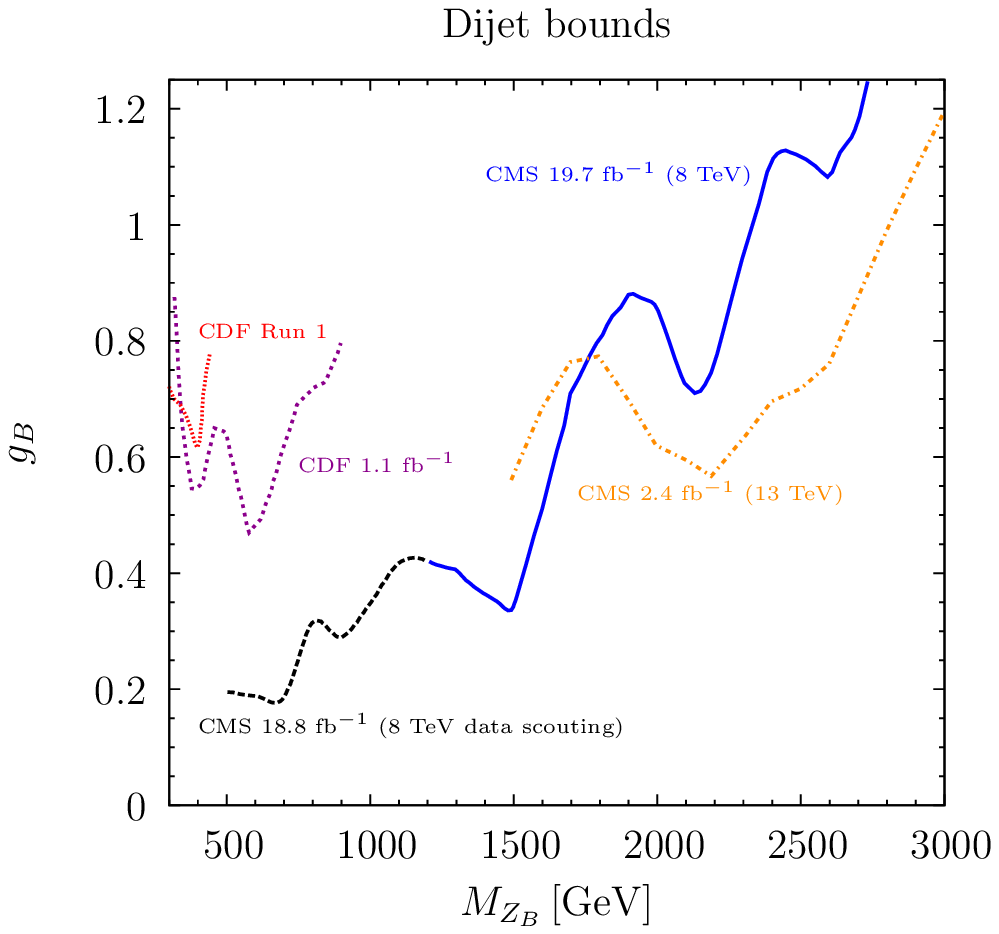}
}
\subfloat[]{
 \includegraphics[width=0.47\linewidth]{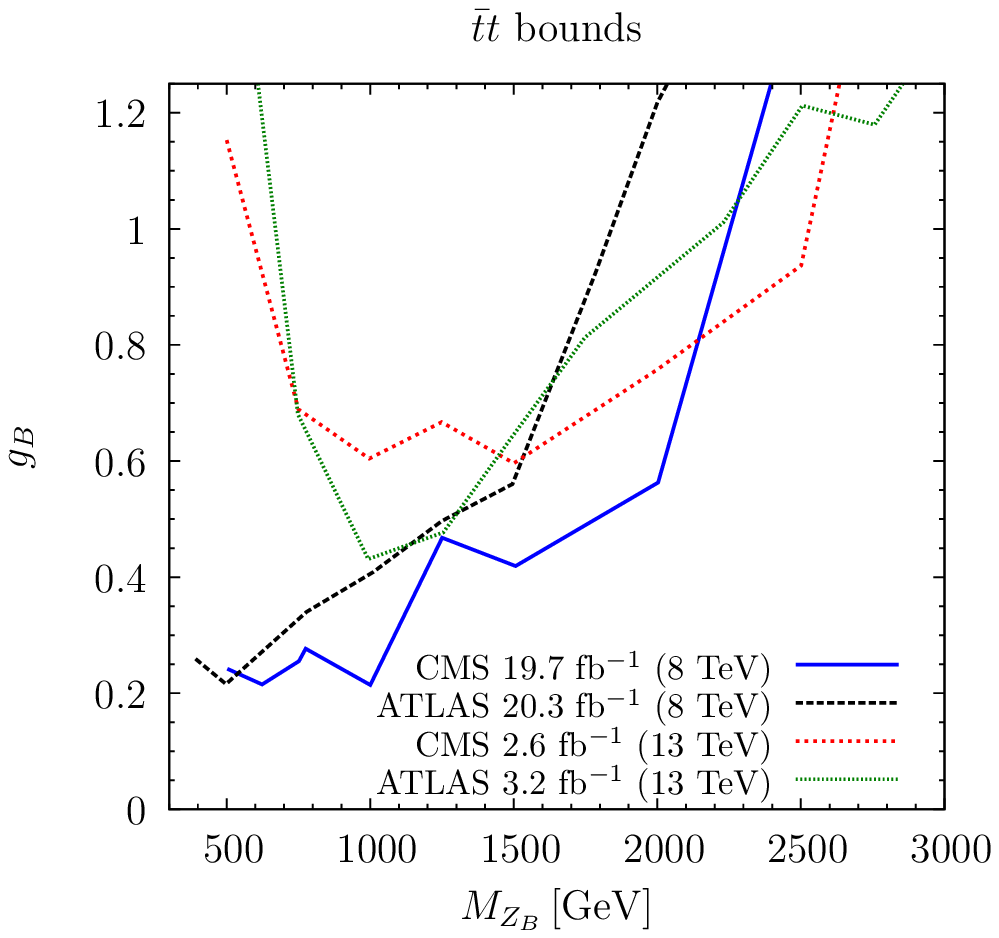}
}
\caption{Limits on the mass and coupling of the leptophobic gauge boson. (a) Limits from dijet searches~\cite{Abe:1997hm,Aaltonen:2008dn,Khachatryan:2015dcf,Khachatryan:2015sja,CMS:2015neg}. (b) Limits from $t\bar{t}$ searches~\cite{Khachatryan:2015sma,CMS:2016zte,ATLAS:2015ttbar,Aad:2015fna}.  \label{fig:gBLimits}}
\end{figure}

\begin{figure}[t]
 \centering
 \includegraphics[width=0.6\linewidth]{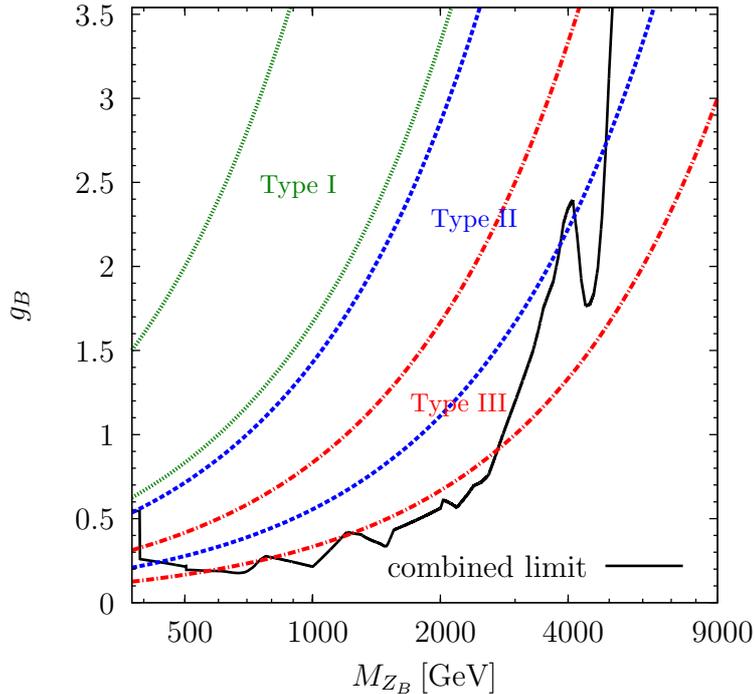}
 \caption{Combined upper limit from dijet and $t\bar{t}$ searches on the gauge coupling $g_B$ and regions that are in agreement with the di-photon excess in the three scenarios discussed in the text: Type I (green dotted), Type II (blue dashed) and Type III (red dash-dotted). \label{fig:allowedRegions}}
\end{figure}

We have shown the possibility to find an upper bound on the symmetry breaking scale and a perturbative upper bound on the leptophobic gauge boson mass.
Now, let us understand if these results are in agreement with the searches for new gauge bosons. In Fig.~\ref{fig:gBLimits} we show the 
experimental bounds in the gauge coupling $g_B$ and mass $M_{Z_B}$ plane from (a) dijet searches and (b) $t \bar{t}$ searches. 
In Fig.~\ref{fig:allowedRegions} we show the combined limit on the gauge coupling from both dijet and $t\bar{t}$ searches, together with the regions that can explain the di-photon excess at the LHC for the three scenarios discussed before. 
As one can see the Type I scenario is ruled out by these experimental constraints. The Type II scenario is highly constrained but there are some 
small regions of the parameter space where one can satisfy the experimental bounds. The Type III scenario is also constrained but the regions which  are close to the upper bound on the symmetry breaking scale are allowed by the experiments.
Therefore, the Type II and Type III scenarios do provide a possible explanation for the di-photon excess in agreement with all experimental constraints.

\begin{table}[b]
\caption{Relative signal strength in the different scenarios. \label{tab:Signalstrength}}
\begin{center}
\begin{tabular}{ccccc}
\hline
~~Scenario~~ & ~~$\Gamma_{W^+ W^-}/\Gamma_{\gamma \gamma}$ ~~& ~~$\Gamma_{ZZ}/\Gamma_{\gamma \gamma}$~~ & ~~$\Gamma_{Z \gamma}/\Gamma_{\gamma \gamma}$~~ & ~~$\Gamma_{g g}/\Gamma_{\gamma \gamma}$~~ \\
\hline \hline
I & 25.4 & 2.3 & 0.5 & 385.5 \\
II & 2.2 & 1.6 & 1.4 & 33.3  \\
III &  0.8 & 0.2 & 0.2 & 11.5 \\
\hline
\end{tabular}
\end{center}
\end{table}   

The different scenarios are experimentally well distinguishable, as the decay rates of the new Higgs boson to other electro-weak gauge bosons and gluons have a different strength. So comparing event rates in the other decay channels with respect to the di-photon event rate provides a powerful testing tool. We give the relative signal strength in Table~\ref{tab:Signalstrength}. It turns out that the most promising signal to distinguish the two viable scenarios II and III  is the measurement of the relative strength of the $Z\gamma$ decay channel, which in the Type II scenario is large enough to search for the channel with two leptons and a photon in the final state.

\section{Summary}
\label{sec:summary}

In this paper we have investigated the possibility to explain the di-photon excess reported by the CMS and ATLAS collaborations in the context of a simple gauge theory.
We have identified the new resonance with a spin zero field which is responsible for the breaking of a new Abelian gauge symmetry. In this context new vector-like quarks 
are needed for an anomaly-free theory and they define the production mechanism and decays of the new Higgs boson. We have focused on a 
simple theory where the local symmetry is baryon number. 

We have shown that if the physical Higgs $h_B$, the field responsible for symmetry breaking, is also responsible for the di-photon excess coming from the $\unit[750]{GeV}$ resonance,
one can find an upper bound on the symmetry breaking scale. Since the masses of the new gauge boson and the vector-like quarks are proportional to the same symmetry breaking 
scale, one can find perturbative upper bounds on their masses. We have investigated three simple scenarios in agreement with cosmology. 
The Type II and Type III scenarios can provide an explanation for the di-photon excess in agreement with the experiments. 
Since we find an upper bound on the symmetry breaking scale, one can hope to discover a new force associated with the local baryon number at the LHC if this theory is relevant for the di-photon excess.

\section*{Acknowledgments}
P.F.P.\ thanks M.\ B.\ Wise for discussions. M.D.\ is supported by the German Science Foundation (DFG) under the Collaborative Research Center (SFB) 676 Particles, Strings and the Early Universe as well as the ERC Starting Grant `NewAve' (638528).

\appendix
\section{Feynman Rules}\label{app:feynman}
 \vspace{-4ex}
\begin{align}
\overline{U_1} U_1 Z &: \ - \frac{i e}{\sin 2 \theta_W} \left( 1 - 2\,Q_u \sin^2 \theta_W \right) \gamma^\mu, \\
 \overline{U_2} U_2 Z &: \ i \,Q_u\, e \tan \theta_W  \ \gamma^\mu, \\
\overline{D_1} D_1 Z &: \ \frac{i e }{\sin 2 \theta_W} \left( 1 + 2 Q_d \sin^2 \theta_W \right) \ \gamma^\mu, \\
\overline{D_2} D_2 Z &: \ i Q_d \, e \tan \theta_W  \ \gamma^\mu, \\
\overline{U_i} U_i A &: \ - Q_u \, i e\, \gamma^\mu, \\
\overline{D_i} D_i A &: \ - Q_d \, i e \gamma^\mu, \\
\overline{U_i} U_i h_B&: \ i \frac{M_{U_i}}{v_B}, \\
\overline{D_i} D_i h_B&: \ i \frac{M_{D_i}}{v_B}.
\end{align}
\section{Decay Widths}
\label{app:decay}
The partial decay widths of the new Higgs $h_B$ are given as follows. In the formulas in this appendix we use $m_f$ and $Q_f$ for the masses and the charges for the vector-like quarks running in the loops. 

\begin{itemize}

\item The decay width into two photons is given by
\begin{align}
\Gamma (h_B \rightarrow \gamma \gamma) = \frac{ \alpha^2}{64 \pi^3} \frac{\left| \sum_f N_c Q_f^2 m_f^2 \left[ 4 M_{h_B}^2 - \left(M_{h_B}^2  - 4 m_f^2 \right) \log^2\left( \frac{\sqrt{1- 4 \frac{m_f^2}{M_{h_B}^2}} - 1}{\sqrt{1- 4 \frac{m_f^2}{M_{h_B}^2}}+1 }\right) \right] \right|^2 }{M_{h_B}^5 v_B^2} .
\end{align}
Here $N_c=3$ is the color factor. For $M_{h_B} < 2 m_f$, this expression can be simplified to
\begin{equation}
 \Gamma (h_B \rightarrow \gamma \gamma) = \frac{ \alpha^2}{4 \pi^3} \frac{\left|  \sum_f N_c Q_f^2 m_f^2 \left[ M_{h_B}^2 + \left( M_{h_B}^2 - 4 m_f^2 \right) \arcsin^2 \left( \frac{M_{h_B}}{2 m_f}\right) \right] \right|^2}{M_{h_B}^5 v_B^2} .
\end{equation}
In the limit $\epsilon= M_{h_B}/m_f \ll 1$ one finds a very simple expression which tells us that there is no decoupling:
\begin{align}
\Gamma (h_B \rightarrow \gamma \gamma)  = \frac{\alpha^2 M_{h_B}^3 \left(\sum_f N_c Q^2_f\right)^2}{144 \pi^3 v_B^2}\,.
\end{align}
\item The decay width into two gluons is given by
\begin{align}
\Gamma (h_B \rightarrow g \, g) = \frac{ \alpha_s^2}{32 \pi^3} \frac{\left| \sum_f m_f^2 \left[ 4 M_{h_B}^2 - \left(M_{h_B}^2  - 4 m_f^2 \right) \log^2\left( \frac{\sqrt{1- 4 \frac{m_f^2}{M_{h_B}^2}} - 1}{\sqrt{1- 4 \frac{m_f^2}{M_{h_B}^2}} +1 }\right) \right] \right|^2 }{M_{h_B}^5 v_B^2}\,.
\end{align}
For $M_{h_B} < 2 m_f$ this expression can be simplified to 
\begin{equation}
 \Gamma (h_B \rightarrow g \, g) = \frac{ \alpha_s^2 }{2 \pi^3} \frac{\left| \sum_f m_f^2 \left[ M_{h_B}^2 + \left( M_{h_B}^2 - 4 m_f^2 \right) \arcsin^2 \left( \frac{M_{h_B}}{2 m_f}\right) \right] \right|^2 }{M_{h_B}^5 v_B^2}\,.
\end{equation}
In leading order in $\epsilon$ this decay width reads for one fermion as
\begin{align}
\Gamma (h_B \rightarrow g\, g) = \frac{\alpha_s^2 M_{h_B}^3}{72 \pi^3 v_B^2}\,.
\end{align}

\item The decay width into $W^+ W^-$ is given by
\begin{align}
\Gamma (h_B \rightarrow W^+ W^-) = g_W^4 \Gamma_V(m_W)\,,
\end{align}
where $g_W = 1/\sqrt{2}$ is the coupling of the new vector-like quarks to the $W$ boson.

\item The decay width into $ZZ$ is given by
\begin{align}
\Gamma (h_B \rightarrow Z\,Z) = \frac{g_Z^4}{2\,c_w^4} \Gamma_V(m_Z)\,,
\end{align}
where $g_Z$ is the coupling of the new vector-like quarks to the $Z$ boson.

\item The general decay width to a vector boson used in the expressions above is (for one fermion) given by
\begin{multline}
\Gamma_V(m_V) = \frac{\alpha^2 N_c^2 m_f^4}{64 \pi^3 m_V^4 M_{h_B} v_B^2 s_w^4 \left( M_{h_B}^2 - 4 m_V^2\right)^2} \sqrt{ 1- \frac{4 m_V^2}{M_{h_B}^2}} \\ \times \left[ \left|A\right|^2 + \left(2 m_V^2 - M_{h_B}^2 \right) \left( A B^\ast + A^\ast B \right) + \left| B \right|^2 \left( M_{h_B}^4 - 4 M_{h_B}^2 m_V^2 + 12 m_V^4 \right) \right],
\end{multline}
where the coefficients are given by
\begin{align}
 A &=   2 \left[M_{h_B}^4 - 6 M_{h_B}^2 m_V^2 + 8 m_V^4 + 2 m_V^2 (M_{h_B}^2 + 2 m_V^2) \left(\Lambda(M_{h_B}^2, m_f, m_f) - \Lambda(m_V^2, m_f, m_f)\right)\right]\nonumber \\ 
 &\quad + (M_{h_B}^2 - 2 m_V^2) \left[ - M_{h_B}^4 + 6 M_{h_B}^2 m_V^2 + 4 m_V^4 + 4 m_f^2 (M_{h_B}^2 - 4 m_V^2)\right] \nonumber \\ &\qquad \qquad \qquad \times C_0(M_{h_B}^2, m_V^2, m_V^2; m_f, m_f, m_f)  ,\\
 B &= 2 \left[ M_{h_B}^2 - 4 m_V^2 + 2 m_V^2 \left(\Lambda(M_{h_B}^2, m_f, m_f) - \Lambda(m_V^2, m_f, m_f)\right)\right] \nonumber \\ 
 &\quad - \left[M_{h_B}^4 - 6 M_{h_B}^2 m_V^2 + 4 m_V^4 - 4 m_f^2 (M_{h_B}^2 - 4 m_V^2)\right] C_0(M_{h_B}^2, m_V^2, m_V^2; m_f, m_f, m_f).
\end{align}
The loop function $\Lambda$ is given by
\begin{align}
 \Lambda(s, m, m) &= \sqrt{1- \frac{4 m^2}{s}} \ln \left( \frac{2 m^2}{2m^2- \left( 1+ \sqrt{1- \frac{4 m^2}{s}}\right) s } \right),
\end{align}
and $C_0$ is the scalar Passarino--Veltman three-point function. 

\item The leading order contribution in the parameter $\epsilon$ of the doublets to the decay to the $W$ bosons is
\begin{align}
\Gamma (h_B \rightarrow W^+ W^-) = \frac{\alpha^2 N_c^2}{72 \pi^3 v_B^2 M_{h_B} s_w^4}\left(M_{h_B}^4  - 4 m_W^2 M_{h_B}^2 + 6 m_W^4 \right) \, \sqrt{ 1- \frac{4 m_W^2}{M_{h_B}^2}} \,.
\end{align}

\item The contribution of the full new quark spectrum expanded to zeroth order in $\epsilon$ to the decay to $Z$ bosons is
\begin{multline}
 \Gamma (h_B \rightarrow Z\,Z) = \frac{\alpha^2 N_c^2 \sqrt{1-\frac{4 m_Z^2}{M_{h_B}^2}}}{576 \pi^3 M_{h_B} s_w^4 v_B^2  c_w^4}  \\ 
 \times \left(M_{h_B}^4-4 M_{h_B}^2 m_Z^2+6
   m_Z^4\right) \left[1 + 2 \left( Q_d - Q_u \right) s_w^2 + 4 \left( Q_d^2 + Q_u^2 \right) s_w^4 \right]^2\,.
\end{multline}

\item The contribution of one new vector-like quark to the decay into $Z$ and $\gamma$ is given by
\begin{align}
\Gamma (h_B \rightarrow Z\,\gamma) = \frac{\alpha^2 m_f^4 Q_f^2 g_Z^2 N_c^2 \,|C|^2}{32 \pi^3\,v_B^2 c_w^2 s_w^2 M_{h_B}\,(m_Z^2 -M_{h_B}^2)^2}\sqrt{1 -\frac{m_Z^2}{M_{h_B}^2}}\,,
\end{align}
with
\begin{multline}
 C = 4 m_Z^2 \left[ \Lambda(m_Z^2; m_f^2, m_f^2) - \Lambda(M_{h_B}^2; m_f^2, m_f^2) \right] + \left(4 m_f^2+ m_Z^2- M_{h_B}^2 \right) \\ 
\times \left[ \log^2\left(\frac{\sqrt{1 -4 \frac{m_f^2}{m_Z^2}} - 1}{\sqrt{1 -4 \frac{m_f^2}{m_Z^2}} + 1} \right)  -   \log^2\left(\frac{\sqrt{1 -4 \frac{m_f^2}{M_{h_B}^2}} - 1}
{\sqrt{1 -4 \frac{m_f^2}{M_{h_B}^2}} + 1} \right) \right] + 4 \left( m_Z^2- M_{h_B}^2 \right) \,.
\end{multline}
The leading order term in $\epsilon$ from the full particle content is given by 
\begin{align}
\Gamma (h_B \rightarrow Z\,\gamma) = \frac{\alpha^2 N_c^2 \sqrt{1-\frac{m_Z^2}{M_{h_B}^2}} \left(m_Z^2- M_{h_B}^2 \right)^2 \left[
 4 \left(Q_d^2+Q_u^2\right)s_w^2+Q_d-Q_u \right]^2}{288 c_w^2
   s_w^2 v_B^2 \pi^3 M_{h_B}}\,.
\end{align}

\item The invisible decay width is given by 
\begin{align}
\Gamma (h_B \rightarrow X \, X) = \frac{\lambda_{BX}^2 v_B^2}{8 \pi M_{h_B}}\sqrt{1 - 4 \,\frac{ m_X^2}{M_{h_B}^2}}\,.
\end{align}

\item The decay width into two new quarks is given by
\begin{align}
\Gamma (h_B \rightarrow \bar{f}\, f ) = \frac{N_c}{8 \pi}  \frac{m_f^2}{v_B^2} M_{h_B} \left(1 - 4 \, \frac{m_f^2}{M_{h_B}^2}\right)^\frac{3}{2}.
\end{align} 
 
\end{itemize}



\begin{thebibliography}{99}

\bibitem{ATLAS:diphotonDec2015}
  The ATLAS collaboration,
  ``Search for resonances decaying to photon pairs in 3.2 fb$^{-1}$ of $pp$ collisions at $\sqrt{s}$ = 13 TeV with the ATLAS detector,''
  ATLAS-CONF-2015-081.

\bibitem{CMS:2015dxe}
  CMS Collaboration [CMS Collaboration],
  ``Search for new physics in high mass diphoton events in proton-proton collisions at 13 TeV,''
  CMS-PAS-EXO-15-004.
  
\bibitem{ATLAS:diphotonMar2016}
  The ATLAS collaboration,
  ``Search for resonances in diphoton events with the ATLAS detector at $\sqrt{s}$ = 13 TeV,''
  ATLAS-CONF-2016-018.
  
\bibitem{CMS:2016owr}
  CMS Collaboration [CMS Collaboration],
  ``Search for new physics in high mass diphoton events in $3.3~\mathrm{fb}^{-1}$ of proton-proton collisions at $\sqrt{s}=13~\mathrm{TeV}$ and combined interpretation of searches at $8~\mathrm{TeV}$ and $13~\mathrm{TeV}$,''
  CMS-PAS-EXO-16-018.
  
\bibitem{Franceschini:2015kwy}
  R.~Franceschini {\it et al.},
  ``What is the $\gamma \gamma$ resonance at 750 GeV?,''
  JHEP {\bf 1603} (2016) 144
  [\href{http://arxiv.org/abs/1512.04933}{arXiv:1512.04933 [hep-ph]}].
  
\bibitem{Ellis:2015oso}
  J.~Ellis, S.~A.~R.~Ellis, J.~Quevillon, V.~Sanz and T.~You,
  ``On the Interpretation of a Possible $\sim 750$ GeV Particle Decaying into $\gamma \gamma$,''
  JHEP {\bf 1603} (2016) 176
  [\href{http://arxiv.org/abs/1512.05327}{arXiv:1512.05327 [hep-ph]}].

\bibitem{Gupta:2015zzs}
  R.~S.~Gupta, S.~Jager, Y.~Kats, G.~Perez and E.~Stamou,
  ``Interpreting a 750 GeV Diphoton Resonance,''
  \href{http://arxiv.org/abs/1512.05332}{arXiv:1512.05332 [hep-ph]}.

\bibitem{McDermott:2015sck}
  S.~D.~McDermott, P.~Meade and H.~Ramani,
  ``Singlet Scalar Resonances and the Diphoton Excess,''
  Phys.\ Lett.\ B {\bf 755} (2016) 353
  [\href{http://arxiv.org/abs/1512.05326}{arXiv:1512.05326 [hep-ph]}].

\bibitem{Dutta:2015wqh}
  B.~Dutta, Y.~Gao, T.~Ghosh, I.~Gogoladze and T.~Li,
  ``Interpretation of the diphoton excess at CMS and ATLAS,''
  Phys.\ Rev.\ D {\bf 93} (2016) 055032
  [\href{http://arxiv.org/abs/1512.05439}{arXiv:1512.05439 [hep-ph]}].
  
\bibitem{Kobakhidze:2015ldh}
  A.~Kobakhidze, F.~Wang, L.~Wu, J.~M.~Yang and M.~Zhang,
  ``750 GeV diphoton resonance in a top and bottom seesaw model,''
  Phys.\ Lett.\ B {\bf 757} (2016) 92
  [\href{http://arxiv.org/abs/1512.05585}{arXiv:1512.05585 [hep-ph]}].

\bibitem{Chao:2015ttq}
  W.~Chao, R.~Huo and J.~H.~Yu,
  ``The Minimal Scalar-Stealth Top Interpretation of the Diphoton Excess,''
  \href{http://arxiv.org/abs/1512.05738}{arXiv:1512.05738 [hep-ph]}.

\bibitem{Curtin:2015jcv}
  D.~Curtin and C.~B.~Verhaaren,
  ``Quirky Explanations for the Diphoton Excess,''
  Phys.\ Rev.\ D {\bf 93} (2016)  055011
  [\href{http://arxiv.org/abs/1512.05753}{arXiv:1512.05753 [hep-ph]}].

\bibitem{Falkowski:2015swt}
  A.~Falkowski, O.~Slone and T.~Volansky,
  ``Phenomenology of a 750 GeV Singlet,''
  JHEP {\bf 1602} (2016) 152
  [\href{http://arxiv.org/abs/1512.05777}{arXiv:1512.05777 [hep-ph]}].

\bibitem{Benbrik:2015fyz}
  R.~Benbrik, C.~H.~Chen and T.~Nomura,
  ``Higgs singlet boson as a diphoton resonance in a vectorlike quark model,''
  Phys.\ Rev.\ D {\bf 93} (2016) 055034
  [\href{http://arxiv.org/abs/1512.06028}{arXiv:1512.06028 [hep-ph]}].

\bibitem{Chao:2015nsm}
  W.~Chao,
  ``Symmetries Behind the 750 GeV Diphoton Excess,''
  \href{http://arxiv.org/abs/1512.06297}{arXiv:1512.06297 [hep-ph]}.

\bibitem{Chang:2015bzc}
  S.~Chang,
  ``A Simple $U(1)$ Gauge Theory Explanation of the Diphoton Excess,''
  Phys.\ Rev.\ D {\bf 93} (2016) 055016
  [\href{http://arxiv.org/abs/1512.06426}{arXiv:1512.06426 [hep-ph]}].

\bibitem{Feng:2015wil}
  T.~F.~Feng, X.~Q.~Li, H.~B.~Zhang and S.~M.~Zhao,
  ``The LHC 750 GeV diphoton excess in supersymmetry with gauged baryon and lepton numbers,''
  \href{http://arxiv.org/abs/1512.06696}{arXiv:1512.06696 [hep-ph]}.

\bibitem{Boucenna:2015pav}
  S.~M.~Boucenna, S.~Morisi and A.~Vicente,
  ``The LHC diphoton resonance from gauge symmetry,''
  \href{http://arxiv.org/abs/1512.06878}{arXiv:1512.06878 [hep-ph]}.

\bibitem{Hernandez:2015ywg}
  A.~E.~C.~Hernandez and I.~Nisandzic,
  ``LHC diphoton 750 GeV resonance as an indication of $SU(3)_c\times SU(3)_L\times U(1)_X$ gauge symmetry,''
  \href{http://arxiv.org/abs/1512.07165}{arXiv:1512.07165 [hep-ph]}.

\bibitem{Pelaggi:2015knk}
  G.~M.~Pelaggi, A.~Strumia and E.~Vigiani,
  ``Trinification can explain the di-photon and di-boson LHC anomalies,''
  JHEP {\bf 1603} (2016) 025
  [\href{http://arxiv.org/abs/1512.07225}{arXiv:1512.07225 [hep-ph]}].

\bibitem{Altmannshofer:2015xfo}
  W.~Altmannshofer, J.~Galloway, S.~Gori, A.~L.~Kagan, A.~Martin and J.~Zupan,
  ``On the 750 GeV di-photon excess,''
  \href{http://arxiv.org/abs/1512.07616}{arXiv:1512.07616 [hep-ph]}.

\bibitem{Cheung:2015cug}
  K.~Cheung, P.~Ko, J.~S.~Lee, J.~Park and P.~Y.~Tseng,
  ``A Higgcision study on the 750 GeV Di-photon Resonance and 125 GeV SM Higgs boson with the Higgs-Singlet Mixing,''
  \href{http://arxiv.org/abs/1512.07853}{arXiv:1512.07853 [hep-ph]}.

\bibitem{An:2015cgp}
  H.~An, C.~Cheung and Y.~Zhang,
  ``Broad Diphotons from Narrow States,''
  \href{http://arxiv.org/abs/1512.08378}{arXiv:1512.08378 [hep-ph]}.

\bibitem{Dev:2015vjd}
  P.~S.~B.~Dev, R.~N.~Mohapatra and Y.~Zhang,
  ``Quark Seesaw, Vectorlike Fermions and Diphoton Excess,''
  JHEP {\bf 1602} (2016) 186
  [\href{http://arxiv.org/abs/1512.08507}{arXiv:1512.08507 [hep-ph]}].
  
\bibitem{Ko:2016lai}
  P.~Ko, Y.~Omura and C.~Yu,
  ``Diphoton Excess at 750 GeV in leptophobic U(1)$^\prime$ model inspired by $E_6$ GUT,''
  \href{http://arxiv.org/abs/1601.00586}{arXiv:1601.00586 [hep-ph]}.

\bibitem{Chao:2016mtn}
  W.~Chao,
  ``The Diphoton Excess from an Exceptional Supersymmetric Standard Model,''
  \href{http://arxiv.org/abs/1601.00633}{arXiv:1601.00633 [hep-ph]}.
  
\bibitem{FileviezPerez:2010gw}
  P.~Fileviez Perez and M.~B.~Wise,
  ``Baryon and lepton number as local gauge symmetries,''
  Phys.\ Rev.\ D {\bf 82} (2010) 011901
  [Erratum: Phys.\ Rev.\ D {\bf 82} (2010) 079901]
  [\href{http://arxiv.org/abs/1002.1754}{arXiv:1002.1754 [hep-ph]}].
  
\bibitem{FileviezPerez:2011pt}
  P.~Fileviez Perez and M.~B.~Wise,
  ``Breaking Local Baryon and Lepton Number at the TeV Scale,''
  JHEP {\bf 1108} (2011) 068
  [\href{http://arxiv.org/abs/1106.0343}{arXiv:1106.0343 [hep-ph]}].

\bibitem{Duerr:2013dza}
  M.~Duerr, P.~Fileviez Perez and M.~B.~Wise,
  ``Gauge Theory for Baryon and Lepton Numbers with Leptoquarks,''
  Phys.\ Rev.\ Lett.\  {\bf 110} (2013) 231801
  [\href{http://arxiv.org/abs/1304.0576}{arXiv:1304.0576 [hep-ph]}].
  
\bibitem{Perez:2014qfa}
  P.~Fileviez Perez, S.~Ohmer and H.~H.~Patel,
  ``Minimal Theory for Lepto-Baryons,''
  Phys.\ Lett.\ B {\bf 735} (2014) 283
  [\href{http://arxiv.org/abs/1403.8029}{arXiv:1403.8029 [hep-ph]}].
  
\bibitem{Perez:2015rza}
  P.~Fileviez Perez,
  ``New Paradigm for Baryon and Lepton Number Violation,''
  Phys.\ Rept.\  {\bf 597} (2015) 1
  [\href{http://arxiv.org/abs/1501.01886}{arXiv:1501.01886 [hep-ph]}].
  
\bibitem{Ohmer:2015lxa}
  S.~Ohmer and H.~H.~Patel,
  ``Leptobaryons as Majorana Dark Matter,''
  Phys.\ Rev.\ D {\bf 92} (2015)  055020
  [\href{http://arxiv.org/abs/1506.00954}{arXiv:1506.00954 [hep-ph]}].
  
\bibitem{Martin:2009iq}
  A.~D.~Martin, W.~J.~Stirling, R.~S.~Thorne and G.~Watt,
  ``Parton distributions for the LHC,''
  Eur.\ Phys.\ J.\ C {\bf 63} (2009) 189
  [\href{http://arxiv.org/abs/0901.0002}{arXiv:0901.0002 [hep-ph]}].
  
\bibitem{Patel:2015tea}
  H.~H.~Patel,
  ``Package-X: A Mathematica package for the analytic calculation of one-loop integrals,''
  Comput.\ Phys.\ Commun.\  {\bf 197} (2015) 276
  [\href{http://arxiv.org/abs/1503.01469}{arXiv:1503.01469 [hep-ph]}].
  
\bibitem{CMS:2015neg}
  CMS Collaboration [CMS Collaboration],
  ``Search for Resonances Decaying to Dijet Final States at $\sqrt{s} = 8$ TeV with Scouting Data,''
  CMS-PAS-EXO-14-005.

\bibitem{Khachatryan:2015dcf}
  V.~Khachatryan {\it et al.} [CMS Collaboration],
  ``Search for narrow resonances decaying to dijets in proton-proton collisions at $\sqrt(s) =$ 13 TeV,''
  Phys.\ Rev.\ Lett.\  {\bf 116} (2016) 071801
  [\href{http://arxiv.org/abs/1512.01224}{arXiv:1512.01224 [hep-ex]}].
  
\bibitem{Khachatryan:2015sja}
  V.~Khachatryan {\it et al.} [CMS Collaboration],
  ``Search for resonances and quantum black holes using dijet mass spectra in proton-proton collisions at $\sqrt{s} =$ 8 TeV,''
  Phys.\ Rev.\ D {\bf 91} (2015) 052009
  [\href{http://arxiv.org/abs/1501.04198}{arXiv:1501.04198 [hep-ex]}].
  
\bibitem{Abe:1997hm}
  F.~Abe {\it et al.} [CDF Collaboration],
  ``Search for new particles decaying to dijets at CDF,''
  Phys.\ Rev.\ D {\bf 55} (1997) 5263
  [\href{http://arxiv.org/abs/hep-ex/9702004}{arXiv:hep-ex/9702004}].
  
\bibitem{Aaltonen:2008dn}
  T.~Aaltonen {\it et al.} [CDF Collaboration],
  ``Search for new particles decaying into dijets in proton-antiproton collisions at s**(1/2) = 1.96-TeV,''
  Phys.\ Rev.\ D {\bf 79} (2009) 112002
  [\href{http://arxiv.org/abs/0812.4036}{arXiv:0812.4036 [hep-ex]}].
  
\bibitem{CMS:2016zte}
  CMS Collaboration [CMS Collaboration],
  ``Search for $\mathrm{t\bar{t}}$ resonances in boosted semileptonic final states in pp collisions at $\sqrt{s}=13~\mathrm{TeV}$,''
  CMS-PAS-B2G-15-002.

\bibitem{Khachatryan:2015sma}
  V.~Khachatryan {\it et al.} [CMS Collaboration],
  ``Search for resonant $t\bar{t}$ production in proton-proton collisions at $\sqrt{s} =$ 8 TeV,''
  Phys.\ Rev.\ D {\bf 93} (2016) 012001
  [\href{http://arxiv.org/abs/1506.03062}{arXiv:1506.03062 [hep-ex]}].

\bibitem{ATLAS:2015ttbar}
  The ATLAS collaboration,
  ``Search for heavy particles decaying to pairs of highly-boosted top quarks using lepton-plus-jets events in proton--proton collisions at $\sqrt{s} = 13$ TeV with the ATLAS detector,''
  ATLAS-CONF-2016-014.

\bibitem{Aad:2015fna}
  G.~Aad {\it et al.} [ATLAS Collaboration],
  ``A search for $ t\overline{t} $ resonances using lepton-plus-jets events in proton-proton collisions at $ \sqrt{s}=8 $ TeV with the ATLAS detector,''
  JHEP {\bf 1508} (2015) 148
  [\href{http://arxiv.org/abs/1505.07018}{arXiv:1505.07018 [hep-ex]}].
  
\end{thebibliography}
\end{document}